\documentclass[aps,twocolumn,secnumarabic,nobalancelastpage,amsmath,amssymb,nofootinbib]{revtex4-2}

\usepackage[utf8]{inputenc}
\usepackage{geometry}
\usepackage{graphicx}
\usepackage{array}
\usepackage{slashed}
\usepackage{amsmath}
\usepackage{amsfonts}
\usepackage{amssymb}
\usepackage[cm]{fullpage}
\usepackage{epsfig}
\usepackage{changes}
\usepackage{comment}
\usepackage{hyperref}
\usepackage{tikz}
\usepackage{feynmp}
\usepackage{cancel}
\usepackage{mathrsfs}
\usepackage{dcolumn}
\usepackage{bm}
\usepackage{balance}
\usepackage{physics}
\usepackage{xspace}
\usepackage{xfrac}
\usepackage{tensor}
\usepackage{mathtools}
\usepackage{ragged2e}
\usepackage{natbib}

\usepackage{floatrow}
\usepackage[acronym]{glossaries}
\glsdisablehyper

\usepackage{cleveref} 
\definechangesauthor[name=Ana,color=violet]{amr}
\definechangesauthor[name=Nuria,color=orange]{nnn}

\DeclareCaptionJustification{justified}{\justifying}

\newcommand*\pFqskip{8mu}
\catcode`,\active
\newcommand*\pFq{\begingroup
        \catcode`\,\active
        \def ,{\mskip\pFqskip\relax}%
        \dopFq
}
\catcode`\,12
\def\dopFq#1#2#3#4#5{%
        {}_{#1}F_{#2}\biggl[\genfrac..{0pt}{}{#3}{#4};#5\biggr]%
        \endgroup
}

\begin{document}

\title{\large{\bf Null quantization, shadows and boost eigenfunctions in Lorentzian AdS}}

\author{Núria Navarro and Ana-Maria Raclariu\vspace{5pt}}

\affiliation{%
  King's College London, Strand, London WC2R 2LS, United Kingdom%
}

\begin{abstract}
We revisit the quantization of a free scalar in 4-dimensional (4d) Lorentzian Anti-de-Sitter spacetime (AdS$_4$). We derive solutions to the wave equation that diagonalize time translations in a foliation of AdS$_4$ with null cones. We show that time-translation eigenmodes of arbitrary mass fields that admit a flat space limit must contain both normalizable and non-normalizable fall-offs as one approaches the boundary along a null leaf. We then show that AdS bulk-to-boundary propagators with suitable time orderings provide alternative bases of solutions to the wave equation. We propose an AdS bulk reconstruction formula relating an on-shell free scalar at a spacetime point to CFT primary operators and their shadow transforms. In the flat space limit, this formula reduces to the Carrollian expansion of a free field in flat space. We finally construct Lorentz boost eigenfunctions in AdS in both hyperbolic and null foliations and show that they respectively become massive and massless conformal primary wavefunctions in the flat space limit.   
\end{abstract}

\maketitle

\nopagebreak

\section{Introduction}

The Anti-de-Sitter/conformal field theory correspondence (AdS/CFT) provides a promising path towards formulating a theory of quantum gravity \cite{Maldacena:1997re,Gubser:1998bc,Witten:1998qj,Aharony:1999ti}. Its successes motivated ongoing efforts to generalize it to theories of gravity closer to our universe, such as asymptotically flat (AF) \cite{deBoer:2003vf,Arcioni:2003xx, Barnich:2006av, Barnich:2010eb,Bagchi:2010zz,Strominger:2013jfa,Duval:2014uva} and de Sitter (dS)  \cite{Banados:1998tb,Lin:1999gf,Strominger:2001pn,Balasubramanian:2001nb,Anninos:2011ui,Arkani-Hamed:2015bza} spacetimes. Progress on unifying approaches towards holography in asymptotically (A)dS and flat space remains rather limited. Our letter takes initial steps in this direction.

Our work is motivated by recent insights into the connection between AdS and flat space holography via a flat space limit \cite{deGioia:2022fcn,deGioia:2023cbd,Bagchi:2023fbj,deGioia:2024yne,Alday:2024yyj,Lipstein:2025jfj,Neuenfeld:2025wnl,deGioia:2025mwt}. While it is long known that CFT correlation functions (or integral transforms thereof) develop S-matrix features in various flat space limits \cite{Giddings:1999qu,Gary:2009mi,Gary:2009ae,Polchinski:1999ry,Penedones:2010ue,Hijano:2019qmi,Hijano:2020szl,Komatsu:2020sag,vanRees:2022zmr}, it was only more recently established that the resulting observables are governed by the infinite dimensional symmetry algebras of AF spacetimes \cite{deGioia:2023cbd,deGioia:2025mwt}. This stems from the fact that conformal algebras in $d > 2$ admit infinite-dimensional enhancements in appropriate limits \cite{Duval:2014uva}. These results were so far obtained by analyzing CFT observables, leaving the relation between the AdS and flat space bulk pictures unclear (\cite{deGioia:2022fcn,Bagchi:2023fbj,deGioia:2024yne,Alday:2024yyj} in fact suggest rather different bulk pictures). Our letter sets the stage for the ambitious goal of identifying the universal bulk origin of these symmetry enhancements in the boundary CFTs. We will focus strictly on  AdS and flat spacetimes, but we expect that our letter will bring useful lessons for dS holography as well. 

We start by revisiting the canonical quantization of a free scalar in Lorentzian AdS$_4$. We first consider a foliation of AdS$_4$ with null cones \cite{Poole:2018koa,Compere:2019bua}. The scalar wave equation in these coordinates admits solutions whose properties differ compared to those in the standard quantization. In this case the two independent solutions to the equation are individually shadow symmetric. Imposing that the wavefunctions are regular at the origin (i.e. have a well-defined flat space limit) leads to wavefunctions that, for arbitrary energies, have both normalizable and non-normalizable fall-offs near the boundary. This behavior is reminiscent of the mixing of source and response near the conformal (null) boundary of AF spacetimes \cite{He:2024vlp} and of principal series representations in dS \cite{Anous:2020nxu}. Indeed, further imposing  normalizability singles out the principal series representations. In AdS, these solutions correspond to imaginary mass fields, but we see no reason why they should be ruled out, as they are a natural output of null quantization. Related observations have appeared in different settings, including Rindler AdS/CFT and the AdS$_2$ near-horizon geometry of near-extremal black holes \cite{Parikh:2012kg, Anninos:2019oka,Sugishita:2022ldv}. 

We next show that bulk-to-boundary propagators provide a basis of solutions to the Klein-Gordon (KG) equation in AdS, allowing for a decomposition of the \textit{bulk} solution space into positive and negative frequency components with respect to the KG inner product.  This is  analogous to the decomposition of a  free field in flat space into plane waves of positive and negative frequencies. To show this, we pick a spacelike slice $\Sigma$ at finite AdS global time and evaluate the KG inner products of bulk-to-boundary propagators with boundary points away from this slice. In this case the non-normalizable asymptotics of the bulk-to-boundary propagators are irrelevant and one obtains (as expected) a result proportional to the standard CFT$_3$ two-point function. The inner product is sensitive to the $i\epsilon$ prescriptions implicit in the definition of the propagators. For instance, inner products of  propagators with the same time ordering differ by a sign depending on whether the boundary points are located above or below $\Sigma$ and vanish for boundary points on opposite sides. On the other hand, the inner products of propagators with opposite time orderings vanish for boundary points on the same side. All products are consistent with Stokes' theorem, conserved away from boundary insertions, and reduce to the inner products of plane waves in both massive and massless flat space limits.

The inner products develop delta function singularities for null separated points. Similar singularities appear in the inner products of a bulk-to-boundary propagator and its shadow as well as in the case where $\Sigma$ is a null surface. We show that these inner products are naturally related to Carrollian and celestial 2-point functions \cite{Raclariu:2021zjz,Pasterski:2021rjz,Bagchi:2025vri}. These results motivated us to propose a new formula \eqref{eq:G-int} expressing a free field in AdS in terms of primary operators in the boundary CFT and their shadow. We discuss some properties and implications of this formula.

Finally, we construct AdS$_4$ wavefunctions that diagonalize a Lorentz $\mathfrak{so}(3,1)$ boost towards a point on the cut defined by the intersection of the bulk null cone through the origin with the boundary. These wavefunctions take different forms depending on whether one uses a hyperbolic foliation (the AdS analog of the Minkowski foliation constructed in \cite{deBoer:2003vf}) or a null foliation. The solutions reduce to the conformal primary wavefunctions of respectively massive and massless particles in the flat space limit. We expect that the null boost wavefunctions are AdS duals to light-ray operators in Lorentzian CFT \cite{Kravchuk:2018htv}. This letter is accompanied by a paper \cite{Navarro:2025} where all  results (and more) are derived in detail.  \vspace{-5pt}\\

{\bf Preliminaries.} Lorentzian AdS$_4$ of radius $\ell$ is the hypersurface
\begin{equation}
    \label{eq:AdS-emb}
  \eta_{AB} X^{A} X^B \equiv  -(X^0)^2 + (\vec{X})^2 - (X^{4})^2 = -\ell^2
\end{equation}
in $\mathbb{R}^{2,3}$. Global AdS$_4$ coordinates are obtained by parameterizing 
\begin{equation}
\label{eq:global-AdS}
\begin{split}
    X^0 &= \ell \frac{\sin \tau}{\cos \rho}, \quad  X^{4} = \ell \frac{\cos\tau}{\cos \rho},\quad
    X^i = \ell \tan \rho \Omega_i.\\  
\end{split}
\end{equation}
Here $\Omega_i$ are unit normals at points on the sphere ($S^{2}$). In these coordinates, the AdS metric takes the form
\begin{equation}
\label{eq:AdS-metric}
ds^2 = \frac{\ell^2}{(\cos \rho)^2}\left( -d\tau^2 + d\rho^2 + \sin^2 \rho d \Omega^2 \right),
\end{equation}
where $d\Omega^2$ is the metric on $S^2$. We will consider the universal cover of Lorentzian AdS, in which case $\tau \in \mathbb{R}$. 

A basis of solutions to the Klein Gordon (KG) equation in AdS is given by the wavefunctions \cite{Penedones:2016voo,Kaplan1}
\begin{equation}
\label{eq:AdS-wf}
\psi_{n\ell J}^{\pm}(\tau, \rho, \Omega) = e^{\mp i\omega_n \tau} f_{n\ell}(\rho) Y_{\ell m}(\Omega).
\end{equation}
Here $Y_{\ell m}$ are spherical harmonics, the radial component takes the form
\begin{equation}
\label{eq:radial}
\begin{split}
f_{n\ell}(\rho) &= \sin^{\ell} \rho \cos^{\Delta} \rho F_{n\ell}(\rho), \\
F_{n\ell}(\rho) &= {}_{2} F_1\left(-n , \ell + \Delta + n;\frac{3}{2} + \ell; \sin^2 \rho \right),
\end{split}
\end{equation}
and $\omega_n$ is quantized
\begin{equation}
    \omega_n \equiv \Delta + 2n + \ell, \quad \Delta(\Delta - 3) = m^2.
\end{equation}
Consequently, an on-shell scalar of mass $m$ in AdS$_4$ can be expressed as
\begin{equation}
\label{eq:AdS-bulk-field}
    \begin{split}
         \Phi(\tau,\rho,\Omega) = \sum_{n, \ell, m}\left( \psi^+_{n \ell m}(\tau,\rho,\Omega) a^{\dagger}_{n \ell m} + h.c. \right).
    \end{split}
\end{equation}
In the quantum theory, $a_{n\ell m}$ and $a^{\dagger}_{n\ell m}$ are promoted to creation and annihilation operators with commutators 
    \begin{equation}
    [a_{n\ell m}, a_{n'\ell' m'}^{\dagger}] \propto \delta_{nn'} \delta_{\ell\ell'} \delta_{mm'}.
\end{equation}
inherited from the orthogonality properties of the wavefunctions \eqref{eq:AdS-wf} with respect to the KG inner product 
\begin{equation}
\label{eq:KG-ip}
\langle \Phi_1, \Phi_2 \rangle \equiv  i \int_{\Sigma} d\Sigma^\mu \left(\Phi_{1}^{\dagger}\partial_{\mu}\Phi_{2}-\Phi_{2}\partial_{\mu}\Phi_{1}^{\dagger}\right).
\end{equation}

\section{Null quantization}
\label{}

We introduce a foliation of AdS$_4$ with null cones obtained by a translation of the null cone through the origin in global time, or equivalently, a rotation in the $(X^0, X^4)$ plane in the embedding space. This yields the parameterization
\begin{equation}
\label{eq:null-foliation}
    X^0 = \cos \tau r + \sin\tau \ell,\quad X^i = r \Omega^i, \quad X^{4} = -\sin\tau r + \cos\tau \ell
\end{equation}
in which the AdS$_{4}$ metric takes the form
\begin{equation}
\label{eq:AdS-retarded}
    ds_{AdS_{4}}^2 = -(\ell^2 + r^2) d\tau^2 - 2\ell d\tau dr + r^2 d\Omega_{2}^2.
\end{equation}
In units of $\ell$, this coincides with the AdS$_4$ metric in Bondi gauge \cite{Poole:2018koa,Compere:2019bua}. 
Solving the KG equation in these coordinates, we find the wavefunctions take the form
\begin{equation}
\label{eq:null-wavefunctions}
    \Psi_{\Delta}(\tau, r, \Omega) = Y_{\ell m}(\Omega ) f(\tau,r), \quad f(\tau, r) = e^{-i\omega \tau} R(r), 
\end{equation}
with  
\begin{equation}
\label{eq:R}
\begin{split}
    R(r) &=  c_1 r^{-1 - \ell} e^{- i\omega \arctan(r)} (1 + r^2)^{\omega/2} \\
    &\times \pFq{2}{1}{\frac{1}{2}(2 - \Delta - \ell + \omega), \frac{1}{2}(\Delta - 1 - \ell + \omega)}{\frac{1}{2} - \ell}{-r^2}  \\
    & + c_2 \times (\ell \leftrightarrow -\ell - 1).
    \end{split}
\end{equation}

Note that each term is individually invariant under the shadow transform $\Delta \rightarrow 3 - \Delta$. This is to be contrasted with the radial part of the wavefunction \eqref{eq:radial} in the global AdS coordinates \eqref{eq:global-AdS}, which is not. This results in different asymptotics near bulk points $r \rightarrow 0$ and the boundary $r \rightarrow \infty$ (see End Matter). Notably, while \eqref{eq:radial} are regular near the origin $\rho \rightarrow 0$ and normalizable for positive $\Delta$, demanding that \eqref{eq:R} are regular near the origin sets $c_1 = 0$ and leads to solutions that, for generic $\omega$, contain both normalizable and non-normalizable fall-offs as $r \rightarrow \infty$. In the special case where $\omega = \Delta + \ell + 2n$, the coefficient of the non-normalizable fall-off vanishes and we recover the standard highest weight representations. 

Conversely, requiring that the solution is normalizable as the boundary is approached along a null cone leads to solutions that are ill-behaved at $r = 0$ for generic $\Delta$ and $\omega$. For $\Delta \in \frac{3}{2} + i\lambda$, corresponding to the principal series representation of $\mathfrak{so}(3,2)$, the wavefunctions are normalizable and admit a bulk-point limit, but the field has imaginary mass. While the wavefunctions \eqref{eq:R} are globally defined, they share some similarities with the Rindler wavefunctions in AdS \cite{Sugishita:2022ldv}, as well as the principal series wavefunctions in $AdS_3/\mathbb{Z}$ constructed in \cite{Melton:2025ecj}. It will be interesting to clarify this connection, as well as the relation to the characteristic initial value problem \cite{Friedrich} in the future.

\section{Scattering bases}

In analogy to \eqref{eq:AdS-wf}, the regular wavefunctions \eqref{eq:null-wavefunctions} should provide continuous positive and negative energy bases of solutions to the scalar wave equation in AdS with $m^2 < -\frac{9}{4}$.
Usually, the boundary limit of the mode expansion of an on-shell field in AdS$_4$ defines an operator in a CFT$_3$. After renormalization, one obtains from \eqref{eq:AdS-bulk-field} a primary operator in a highest-weight representation of $\mathfrak{so}(3,2)$ leading to the extrapolate AdS/CFT dictionary in the simplest case of a free scalar \cite{Banks:1998dd}. It is less clear how one could apply the same reasoning to an expansion of the field in the basis constructed from \eqref{eq:R}. Either way, neither the GKPW \cite{Gubser:1998bc,Witten:1998qj} nor the extrapolate \cite{Banks:1998dd,Harlow:2011ke} dictionaries provide a map that allows for an on-shell scalar at \textit{any} bulk point to be expressed in terms of operators in the boundary CFT.
The HKLL reconstruction \cite{Hamilton:2005ju,Hamilton:2006fh,Hamilton:2006az} and its extensions \cite{Dong:2016eik,Harlow:2018fse} proposed a partial solution to this problem, which we revisit next.

\subsection{Bulk-boundary map, revisited}

In the HKLL proposal, fields in AdS are ``reconstructed'' by integrating CFT primary operators over a boundary region consisting of points spacelike separated from the bulk point. As a result, one has to enlarge the boundary integration region to access points deeper into the AdS bulk. The reconstruction of regions beyond the causal horizon remains rather challenging in practice \cite{Papadodimas:2012aq,Almheiri:2014lwa,Dong:2016eik,Chen:2019gbt,Leutheusser:2022bgi, Caron-Huot:2025she,Caron-Huot:2025hmk}. One also runs into problems when considering high-energy scattering in AdS, where we would expect smearing over infinitesimal strips of the boundary to be sufficient (at least in a high-energy limit) to access the region near the bulk point \footnote{It was nevertheless shown in \cite{Hijano:2019qmi,Hijano:2020szl} that the HKLL reconstruction involving smearing over the full boundary localizes to strips separated by $\Delta \tau = \pi$ in the flat space limit.}. This situation is in sharp contrast with Minkowski space, where bulk fields (at least in free theories) can in fact be reconstructed from their value at \textit{cuts} of $\mathscr{I}^{\pm}$ \cite{cite-key}. 

In this section we propose a general formula that allows for a free scalar in AdS$_4$ to be expressed in terms of operators in the boundary CFT$_3$ schematically as
\begin{equation}
    \label{eq:G-int}
    \begin{split}
    \Phi(\tau, \rho, \Omega) &= \alpha \int_{B} d\tau' d\Omega' {\bf G}_{3 - \Delta}(\tau', \Omega';\tau, \rho, \Omega) \mathcal{O}_{\Delta}(\tau', \Omega') \\
    &+ \beta \int_{B} d\tau' d\Omega' {\bf G}_{\Delta}(\tau', \Omega';\tau, \rho, \Omega) \widetilde{\mathcal{O}}_{3 - \Delta}(\tau', \Omega') .
    \end{split}
\end{equation}
Here ${\bf G}_{\Delta}$ and ${\bf G}_{3 - \Delta}$ consist of AdS bulk-to-boundary propagators \cite{Witten:1998qj,Sleight:2023ojm, H:2024cfo} with specified time orderings, $\mathcal{O}_{\Delta}$ and $\widetilde{\mathcal{O}}_{3-\Delta}$ are CFT$_3$ primaries of dimensions $\Delta$ and $3 - \Delta$, $B$ is a boundary region and $\alpha, \beta$ are determined by demanding that the field has normalizable fall-off near the boundary. As we will show in our companion paper \cite{Navarro:2025}, 
the asymptotics of bulk-to-boundary propagators differ depending on whether one approaches the boundary along a spacelike or a null direction. $\alpha$ and $\beta$ will consequently depend on this choice. Furthermore, we show that \eqref{eq:G-int} is in agreement with the extrapolate dictionary provided that $\widetilde{\mathcal{O}}_{3 -\Delta}$ and $\mathcal{O}_{\Delta}$ are related by a (Lorentzian version of the) shadow transform \cite{Ferrara:1972uq,Dolan:2003hv,Simmons-Duffin:2012juh}. \eqref{eq:G-int} has the potential of unifying distinct proposals in (A)dS and flat holography. Indeed, a similar formula was proposed in the dS/CFT context in \cite{Xiao:2014uea}, and shadow operators were also argued to play a role in the construction of celestial amplitudes \cite{Kapec:2016jld,Pasterski:2017kqt,Crawley:2021ivb,Banerjee:2024yir} and the bootstrap of cosmological correlators \cite{Sleight:2019hfp}.

\subsection{A basis of bulk-to-boundary propagators}
\label{sec:in-out}

Eq. \eqref{eq:G-int} is motivated by the observation that bulk-to-boundary propagators can be used to construct bases of solutions to the AdS wave equation with respect to the AdS inner product \eqref{eq:KG-ip}. 
 This turns out to \textit{not} be in tension with the well-known fact that $G_{\Delta}$ has both normalizable and non-normalizable fall-offs as one approaches the boundary along a spacelike direction \cite{Witten:1998qj}. 
 
 The key to establishing this claim is the inner product of bulk-to-boundary propagators on a spacelike slice in AdS. Surprisingly, we could not find a computation of this elementary object in the literature. Adjacent results can be found in \cite{Costa:2014kfa, Melton:2025ecj} (see also \cite{Berenstein:2025tts} where this gap has been recently pointed out and partially addressed). This inner product can be computed using the decomposition of retarded, advanced and (anti)-time-ordered bulk-to-boundary propagators in the \eqref{eq:AdS-wf} basis \cite{Giddings:1999qu} -- see \cite{Navarro:2025}. Given the normalized time-ordered bulk-to-boundary propagator
\begin{equation}
\label{eq:TO}
    G_{\Delta}(P;X) \equiv \frac{2^{\Delta - 1}\Gamma(\Delta)}{\pi^{3/2}\Gamma(\Delta - \frac{1}{2})}\frac{1}{(-P\cdot X + i\epsilon)^{\Delta}},
\end{equation}
taking $\Sigma$ to be a slice of constant time $\tau = \tau_0$ and the boundary insertions \textit{away} from the slice, we find
 \begin{equation}
    \label{eq:time-ordered-btb-ip} 
\begin{split}
\langle G_{\Delta}(P_1; X), &G_{\Delta}(P_2; X) \rangle_{\tau = \tau_0} \\
= 2^{2\Delta} &\left( G_{\Delta}(P_2, P_1) \Theta(\tau_{01}) \Theta(\tau_{02}) \Theta(\tau_{12}) \right. \\
&\left. - G_{\Delta}(P_2, P_1) \Theta(\tau_{10}) \Theta(\tau_{20})\Theta(\tau_{21})\right).
\end{split}
\end{equation}
The inner product of anti-time-ordered propagators is obtained by complex conjugation.

Eq. \eqref{eq:time-ordered-btb-ip} is non-vanishing only for boundary points located either above or below the slice on which the inner product is evaluated. In contrast, the inner product of a time-ordered and an anti-time-ordered propagator turns out to be non-vanishing only for boundary points located on opposite sides of the slice. The inner products agree (up to sign) with the CFT$_3$ two-point function upon symmetrization under $P_1 \leftrightarrow P_2$. 
We conclude that the (anti)-time-ordered bulk-to-boundary propagators with the boundary points inserted (above) below $\Sigma$ form a positive frequency basis with respect to the KG inner product evaluated on $\Sigma$. One can show that this basis is orthogonal and delta-function normalizable with respect to a \textit{shadow} KG inner product \cite{Crawley:2021ivb} (see also \eqref{eq:shadow-G}). 

\begin{figure}
   \centering
   \includegraphics[width=0.6\linewidth]{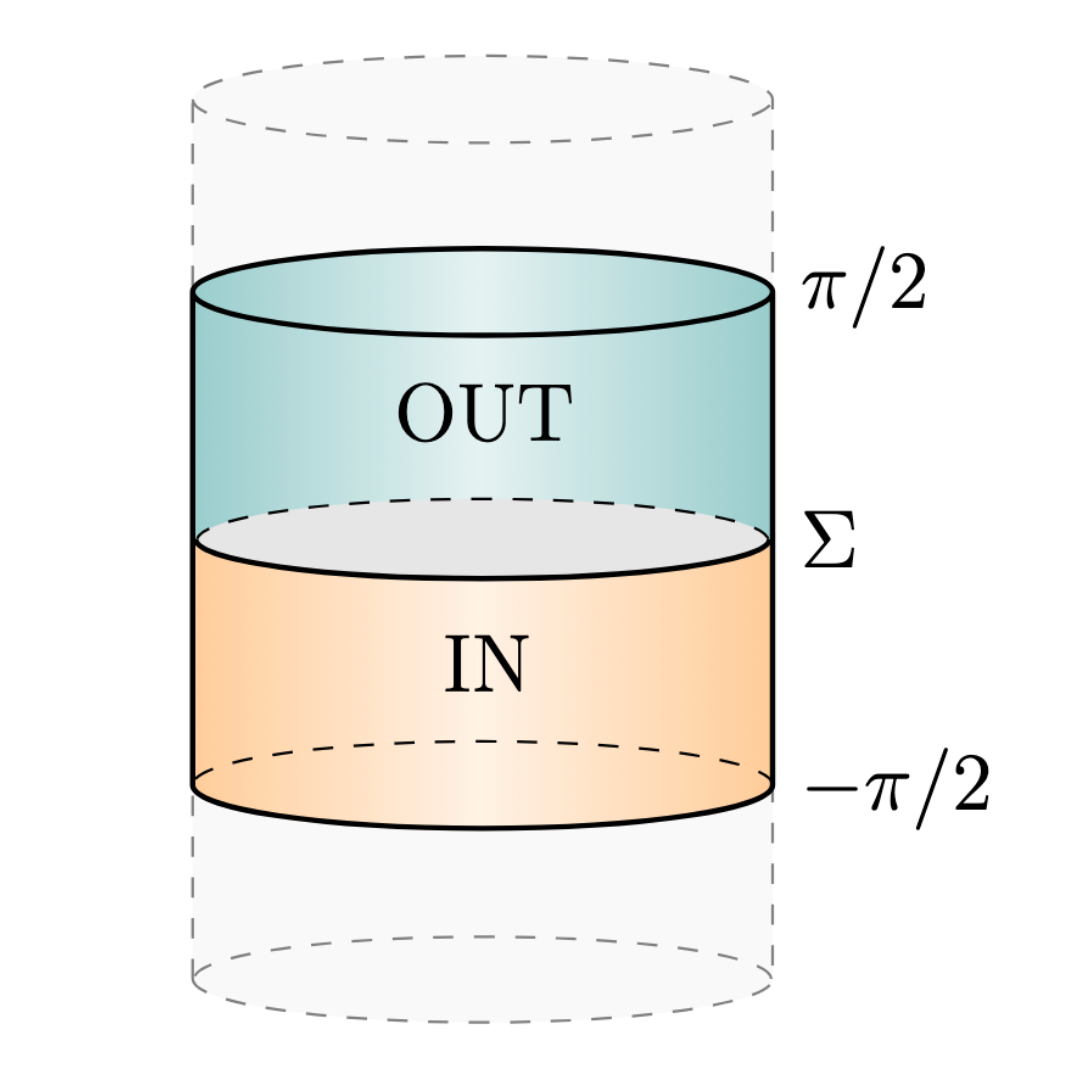}
   \caption{Time-ordered bulk-to-boundary propagators provide a decomposition of the space of solutions to the KG equation into positive/negative frequency components for boundary points in the in/out regions. The in/out sectors are orthogonal with respect to the KG inner product on $\Sigma$, in analogy with the plane wave decomposition of solutions to the KG equation in flat space.}
    \label{fig:in-out}
\end{figure}

We will refer to the positive and negative frequency components spanned by $G_{\Delta}$ as in/out bases in AdS, as their properties are identical to those obeyed by incoming and outgoing plane waves in Minkowski space. One can also show (see \cite{Navarro:2025}) that the KG inner products such as \eqref{eq:time-ordered-btb-ip} reduce to those of plane waves in either massive or massless flat space limits. 
We illustrate this in Figure \ref{fig:in-out}. A similar picture was proposed in \cite{Alday:2024yyj} (see Figure 5 therein) as relating AdS correlators to flat space amplitudes in a flat space limit. Here we see that the flat space limit or the Poincar\'e patches are not crucial ingredients in the construction. In particular, different choices of (spacelike) $\Sigma$ will yield the same decomposition of the Hilbert space provided that no boundary insertions are crossed. We next show that novel features appear in the case where $\Sigma$ is null.  

\subsection{Shadow basis vs. inner product on null $\Sigma$}

Eq. \eqref{eq:G-int} also involves ${\bf G}_{3 - \Delta}$, consisting of components related to $G_{\Delta}$  by a shadow transform. 
Shadow transforms have been discussed extensively in the bootstrap of Euclidean CFTs \cite{Ferrara:1972uq,Dolan:2003hv,Simmons-Duffin:2012juh}, as well as celestial and dS holography \cite{Kapec:2016jld,Sleight:2019hfp,Crawley:2021ivb,Pasterski:2021fjn,Banerjee:2022wht,Himwich:2025bza}, but less so in the  context of Lorentzian AdS/CFT. We find that the $i\epsilon$ prescriptions in the shadow Kernel must be correlated with the time-ordering of the propagator \cite{Navarro:2025}. In this case, the shadow transform maps bulk-to-boundary propagators with boundary points inserted at times before $X$ to ones with boundary points at times later than $X$. This is a reflection of the fact that the shadow transform is related to inversions in a CFT \cite{Chen:2023tvj,Jorstad:2023ajr,Chen:2024kuq}.  
 Upon symmetrization $P_1 \leftrightarrow P_2$, the inner product of the time-ordered $G_{\Delta}(P_1,X)$ and $G_{3 - \Delta}(P_2,X)$ on $\Sigma$ takes the form
\begin{equation}
\label{eq:shadow-G}
    \begin{split}
&{\rm Sym}_{1 \leftrightarrow 2}\langle G_{\Delta}^{}(P_1;X), G_{3 - \Delta}^{}(P_2;X)\rangle_{\tau = \tau_0} \propto i \delta(\tau_{12}+k\pi)\\
&\times \left[  \Theta(\tau_{01})\Theta(\tau_{02}) - \Theta(\tau_{10})\Theta(\tau_{20})\right]  \delta^{(2)}(\Omega_1 - \Omega_2^k).
\end{split}
\end{equation}
Here $\Omega_2^k \equiv (-)^{k}\Omega_2$.  
The inner product of time-ordered and anti-time-ordered propagators is similar, but non-vanishing only if the boundary points are on opposite sides of $\Sigma$. Interestingly, this implies that in the case where either the in or the out bases of Section \ref{sec:in-out} are replaced with $G_{3 - \Delta}$ with the appropriate time ordering, the in-out two-point functions become delta functions that are only non-vanishing for null-separated boundary points. This is precisely the behavior of a two-point functions in celestial and Carrollian CFTs inherited from the on-shell propagator in Mink$_4$ \cite{Pasterski:2017kqt,Donnay:2022wvx}.   

It is also interesting to compare \eqref{eq:time-ordered-btb-ip} to the inner product evaluated on a null surface in AdS. For boundary points on the null plane, we find \cite{Navarro:2025}
\begin{equation}
\label{eq:null}
\begin{split}
    \langle G_{\Delta}^{}(P_1, X), G_{\Delta}^{}(P_2, X)\rangle_{\rm null} 
    &\propto \epsilon^{2 - 2\Delta} \delta^{(2)}(w_1 - w_2).
    \end{split}
\end{equation}
For boundary points away from the null slice, we expect to recover \eqref{eq:time-ordered-btb-ip}.
Up to normalization, this inner product coincides with the time-ordered -- anti-time-ordered counterpart of \eqref{eq:shadow-G}. The normalization precisely agrees with the one relating CFT$_3$ and Carrollian/celestial primaries \cite{deGioia:2022fcn,deGioia:2024yne,Alday:2024yyj}. 

We also note that the bulk-to-boundary propagators have different near-boundary asymptotics along $\Sigma$. Using the null coordinates \eqref{eq:null-foliation} and letting $r, r_p \rightarrow \infty$ for $\frac{r}{r_p} = 1$, we find 
\begin{equation}
\label{eq:G-null}
    \lim_{r \rightarrow \infty} r^{\Delta} G_{\Delta}(P;X) = \frac{1}{\left( \cos(\tau - \tau_p) - \Omega \cdot \Omega_p +i\epsilon \right)^{\Delta}},
\end{equation}
where the subscript $p$ refers to points in the boundary.
The near-boundary expansion of $G_{\Delta}$ along a null cone therefore only contains the $\sim r^{-\Delta}$ fall-off. Note that the near-boundary asymptotics of bulk-to-boundary propagators for different foliation choices is precisely opposite to those of the corresponding wavefunctions \eqref{eq:AdS-wf}, \eqref{eq:null-wavefunctions}.  Consequently, demanding that $\Phi$ in \eqref{eq:G-int} is normalizable as one approaches the boundary along a spacelike slice implies that $\alpha \propto \beta \neq 0$, while the same requirement for the null slice implies that $\alpha = 0, \beta \neq 0$ in \eqref{eq:G-int}. In our companion paper \cite{Navarro:2025} we show that the former case is in complete agreement with the extrapolate dictionary, while the latter choice clarifies the holographic dictionary in flat space and its relation to the flat space limit of AdS/CFT, as we summarize next.

\subsection{Relation to flat space holography}

It has recently been shown that CFT$_3$ correlators develop flat space features in various related limits \cite{Hijano:2020szl,deGioia:2022fcn,Bagchi:2023fbj,deGioia:2023cbd, Alday:2024yyj,deGioia:2024yne,Lipstein:2025jfj,Kulkarni:2025qcx,Surubaru:2025fmg,Fontanella:2025tbs}. In this section, we clarify the bulk picture and also derive a novel relation between shadow operators in CFT$_3$ and Carrollian/celestial operators. The CFT$_3$ shadow featured in the derivation \cite{deGioia:2023cbd} of the celestial CFT supertranslation and superrotation Ward identities \cite{He:2014laa,Kapec:2016jld} from the CFT stress tensor Ward identities, but the reason for its appearance remained unexplained. Imposing that the AdS field is finite as the boundary is approached along a null direction sets $\alpha = 0$ in \eqref{eq:G-int}. The bulk-point limit of CFT correlators further suggests that we should consider a boundary integration region $B$ 
\begin{equation}
    \label{eq:boundary-retarded}
 \tau_p = \tau + c (u - u_p), \quad u_p \in [u_i, u_f], \quad c = \frac{1}{\ell}.
\end{equation}
Then taking $\ell \rightarrow \infty$ keeping all other coordinates and $\Delta$ fixed, we obtain
\begin{equation}
\begin{split}
\Phi(X) &= c  \int_{u_i}^{u_f}  du_p\int d\Omega_p \frac{1}{(-(u - u_p) - r\Omega \cdot \Omega_p + r + i\epsilon)^{\Delta}} \\
&\times {\widetilde{\mathcal{O}}}_{3 - \Delta}(\tau + c(u - u_p), \Omega_p)
\end{split}
\end{equation}
which agrees with the Carrollian expansion of a scalar field in flat space \cite{Bagchi:2023fbj} provided that the Carrollian operator is identified with the \textit{shadow transform} of a CFT$_3$ operator. 

This relation is substantiated by the fact that the inner product of the AdS bulk-to-boundary propagator and its shadow is a solution to the conformal Carroll Ward identity. Such solutions take the form \cite{Bagchi:2025vri}
\begin{equation}
\label{eq:Carr-two-point}
\begin{split}
    &\langle O_{\Delta_1}(u_1, z_1) O_{\Delta_2}(u_2, z_2)\rangle \propto f(u_{12}) \delta^{(2)}(z_{12}), \\
    & f(u_{12}) = \frac{1}{(u_{12} + i\epsilon)^{\Delta_1 + \Delta_2 - 2}}.
    \end{split}
\end{equation}
For $\Delta_2 = 3 - \Delta_1$, this agrees with the inner product \eqref{eq:shadow-G} prior to symmetrization. Since \eqref{eq:Carr-two-point} is a solution for either $i\epsilon$ prescription, so is $f(u_{12}) \propto \delta(u_{12})$. Similarly, the null KG inner product \eqref{eq:null} can be seen to take the same form in \eqref{eq:Carr-two-point} with $\Delta_1 = \Delta_2  = \Delta$ and $u_{12} \rightarrow \epsilon$. It will be interesting to explore the relation between the geometric quantization perspective employed here and the path integral \cite{Kim:2023qbl,Jain:2023fxc}.

\section{Boost eigenstates in AdS}
\label{sec:boosts}

To further elucidate the tight connections between AdS and flat holography, in this section we derive solutions to the scalar wave equation in Lorentzian AdS$_4$ that diagonalize boosts towards a cut of the conformal boundary. The construction is precisely analogous to that of conformal primary wavefunctions in Mink$_4$ \cite{deBoer:2003vf,Pasterski:2016qvg,Pasterski:2017kqt}. We will see that these solutions take different forms depending on whether one chooses to foliate spacetime with hyperbolic slices (mimicking the construction in \cite{deBoer:2003vf}) or null cones. We expect these wavefunctions to span different irreducible representations of the $\mathfrak{so}(3,1)$ sublagebra of $\mathfrak{so}(3,2)$. As a result, we obtain a decomposition of an AdS scalar particle (ie. a representation of dimension $\Delta$) into irreps of its $\mathfrak{so}(3,1)$ subalgebra. 

\subsection{Hyperbolic foliation}

\begin{figure}
    \centering
    \includegraphics[width=\linewidth]{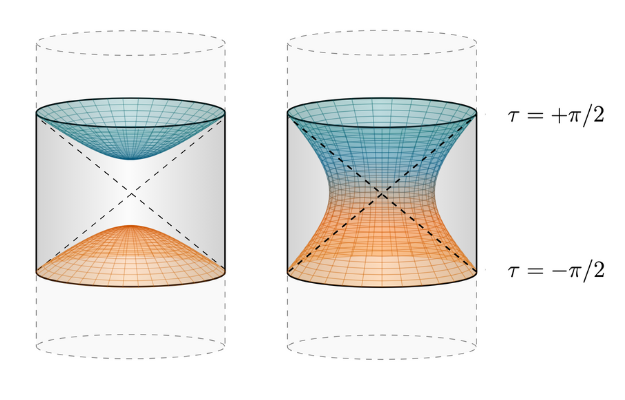}
    \caption{Schematic representation of AdS$_4$ foliated into EAdS$_3$ (left) and dS$_3$ (right) leaves.}
    \label{fig:foliations}
\end{figure}

Consider the null cone through the origin of AdS$_4$. We obtain a hyperbolic foliation AdS$_4$ by setting
\begin{equation}
\label{eq:AdSd-emb}
     -(X^0)^2 + \sum_{i = 1}^3 (X^i)^2 = -\alpha^2, \quad X^{4} = \pm \sqrt{\ell^2 - \alpha^2}.
\end{equation}
In this parameterization, the AdS$_4$ metric takes the form
\begin{equation}
\label{eq:massive-AdS-fol}
    ds^2_{4} = -\ell^2 \frac{d\alpha^2}{\ell^2 - \alpha^2} + \alpha^2 ds_{3}^2,
\end{equation}
where $ds_3^2$ is the metric on the Euclidean (E)AdS$_3$ leaves of radii $\alpha$ in the future/past of the lightcone (Milne regions). The exterior of the lightcone is instead foliated with dS$_3$ leaves, obtained by $\alpha \rightarrow i\alpha$. $\alpha = 0$ corresponds to the null cone through the origin, while $\alpha = \pm \ell$ correspond to the hypersurfaces of radius $\ell$ at $\tau = \pm \frac{\pi}{2}$. 
We illustrate this in Figure \ref{fig:foliations}. 
Note that in the $\ell \rightarrow \infty$ limit, \eqref{eq:massive-AdS-fol} reduces to the Mink$_4$ foliation introduced in \cite{deBoer:2003vf}. 

In the coordinates \eqref{eq:massive-AdS-fol}, the AdS$_4$ Laplacian in units of $\ell$ takes the form
    \begin{equation}
    \Box_{AdS_{4}}  = \alpha^{-2} \Box_{AdS_{3}} + \alpha^{-1}\left((4\alpha^2 - 3) \partial_{\alpha} + \alpha (\alpha^2 -1) \partial_{\alpha}^2 \right).
\end{equation}
The solutions to the KG equation can be expressed in terms of the wavefunctions
\begin{equation}
    \label{eq:AdS4-cpw}
    \Psi_{\Delta}(\delta,\hat{q}; \alpha,x) = f(\alpha) G_{AdS_{3}}^{\delta}(\hat{q};x),
\end{equation}
where $\hat{q}, x \in \mathbb{R}^{3,1}$ with $\hat{q}^2 = 0,~ x^2 = -1$, $G_{AdS_3}^{\delta}(\hat{q};x)$ is the bulk-to-boundary propagator on EAdS$_3$, $\delta$ is a boost weight and $f(\alpha)$ satisfies the ODE
\begin{equation}
\label{eq:massive-f}
\begin{split}
    \alpha \left((4\alpha^2 - 3) \frac{d}{d\alpha} \right. &+\left. \alpha (\alpha^2 -1) \frac{d^2}{d\alpha^2} \right)f(\alpha) \\
  &= \left( \alpha^2\Delta(\Delta - 3) - \delta(\delta - 2) \right) f(\alpha).
  \end{split}
\end{equation}
 Solutions to this equation take the form
\begin{equation}
\label{eq:f}
\begin{split}
 f(\alpha) &= c_1  \alpha^{-\delta} {}_2F_1\left(\frac{1}{2}(3 - \delta - \Delta), \frac{1}{2}(\Delta - \delta); 2 - \delta ;\alpha^2 \right) \\
 &+ c_2 \times (\delta \leftrightarrow 2 - \delta).
 \end{split}
    \end{equation}
By construction, \eqref{eq:AdS4-cpw} diagonalize the AdS$_4$ boosts towards the point $\hat{q}$ on its conformal boundary. 

It can be shown using Sturm-Liouville theory that $f(\alpha)$ are orthogonal for $\alpha \in [0,1)$ with respect to the inner product
\begin{equation}
    \langle f, g \rangle = \int_0^{1} d\alpha \frac{\alpha}{\sqrt{1 - \alpha^2}} f(\alpha) g(\alpha)
\end{equation}
provided that $\delta \in 1 + i\lambda$, $\lambda \in \mathbb{R}$. Note that for fixed $\lambda,$ $G_{AdS_3}^{\delta}(\hat{q};x)$ span a principal series representation of $\mathfrak{so}(3,1)$ \cite{Costa:2014kfa,Pasterski:2017kqt}. Using orthogonality of bulk-to-boundary propagators, we conclude that solutions to the KG equation in the future Milne patch can be expressed as
\begin{equation}
    \Psi_{\Delta}(X(\alpha,x)) = \int_{\mathbb{R}} d\lambda \int d^{2} \hat{q}  f(\alpha)  G^{1 + i\lambda}_{AdS_3}(\hat{q};x) \mathcal{O}_{1 - i\lambda}(\hat{q}).
\end{equation}
Adding up the contributions from the different E(A)dS regions, we obtain the AdS$_4$ analog of the decomposition of a massive solution to the Mink$_4$ wave equation in terms of principal series representations of $\mathfrak{so}(3,1)$ \cite{deBoer:2003vf,Pasterski:2016qvg,Pasterski:2017kqt}. 

The analogy can be made clearer by writing the wavefunctions \eqref{eq:AdS4-cpw} (away from the lightcone) as
\begin{equation}
\label{eq:hyper-int-rep}
     \Psi_{\Delta}(\delta, \hat{q}; \alpha, x) = N \int_{AdS_3} [dx'] G^{\delta}_{AdS_3}(x'; \hat{q}) (-\alpha x \cdot x' + i\epsilon)^{-\Delta},
\end{equation}
where the integral is evaluated over the EAdS$_3$ leaf of radius $\ell$. We show this explicitly in our companion paper \cite{Navarro:2025}. $(-\alpha x\cdot x' + i\epsilon)^{-\Delta}$ here play the same role as the massive plane waves in flat space. \vspace{-8pt}\\

{\bf Flat space limit.} We now show that the wavefunctions \eqref{eq:AdS4-cpw} become \textit{massive} conformal primary wavefunctions in the limit $\Delta \sim \ell \rightarrow \infty$. We only need to show that \eqref{eq:f} becomes a Bessel function of the second kind in the limit, namely \cite{Pasterski:2016qvg,Pasterski:2017kqt,Iacobacci:2022yjo,Sleight:2023ojm}
\begin{equation}
    \lim_{\ell \sim \Delta \rightarrow \infty} f(\alpha) \propto \alpha^{-1} K_{\delta - 1}(i m \alpha).
\end{equation}
Restoring the factors of $\ell$, we find
\begin{equation}
     \begin{split}
   \lim_{\ell \sim \Delta \rightarrow \infty} &{}_2F_1\left(\frac{\delta +1 - \Delta}{2}, \frac{\delta - 2 + \Delta}{2}, \delta, \frac{\alpha^2}{\ell^2} \right) \\
   &= \Gamma(\delta)\left(\frac{2}{i\alpha} \right)^{\delta-1}I_{\delta-1}(i\alpha),
   \end{split}
\end{equation}
where $I_{\delta - 1}$ is the modified Bessel function of first kind. 
The relation between Bessel functions of first and second kind then determines the ratio of coefficients in \eqref{eq:f} 
\begin{equation}
    \frac{c_1 4^{\delta-1} \Gamma(\delta)}{c_2  \Gamma(2 - \delta)} = -1.
\end{equation}

\subsection{Null foliation}

In flat space, the massless conformal primary wavefunctions can be obtained by taking the massless limit of the analog of \eqref{eq:hyper-int-rep}. We can do the same in AdS by introducing Poincar\'e coordinates on EAdS$_3$ $x' = x'(y,z,\bar{z})$ and taking the limit $y, \alpha \rightarrow 0$ for fixed
   $ r \equiv \frac{\alpha}{y}. $
This limit is singular, and while it does lead to boost eigenstates of dimension $\delta$, the resulting wavefunctions  fail to obey the AdS$_4$ wave equation. To obtain boost eigenfunctions that solve the wave equation, we replace $(-\alpha x \cdot x')^{-\Delta}$ by the bulk-to-bulk propagator in AdS$_4$ \cite{Burgess:1984ti}. Then, parameterizing the bulk point $X$ with the null coordinates \eqref{eq:null-foliation}, we find
\begin{equation}
\label{eq:final-null-sol}
\begin{split}
    &\Psi_{\Delta}^{\rm null}(\delta, \hat{q}(z);\tau, r, w) 
    = \frac{\Gamma(\Delta - \delta)}{\Gamma(\Delta - \frac{1}{2})}  \\
    & \times \pFq{2}{1}{\frac{1}{2}(\Delta - \delta), \frac{1}{2}(\Delta - \delta + 1)}{\Delta - \frac{1}{2}}{\frac{1}{(\ell \cos\tau - r \sin \tau)^2}}\\
   & \times \frac{(\ell^2 \cos \tau - \ell r \sin \tau )^{\delta-\Delta}}{(r \cos \tau + \ell \sin \tau - r \Omega \cdot \Omega_q)^{\delta}} .\\
    \end{split}
\end{equation}
It can be checked using the definition of the hypergeometric function that \eqref{eq:final-null-sol} are solutions to the KG equation in AdS$_4$ that diagonalize boosts towards $\hat{q}$ (see \cite{Navarro:2025} for details). \vspace{-8pt}\\

{\bf Flat space limit.} We now consider the flat space limit $\tau \rightarrow 0, \ell \rightarrow \infty$ for fixed $\tau \ell \equiv u$ and $\Delta$. In this limit, the hypergeometric function function is set to 1 and we recover the conformal primary wavefunction of a \textit{massless} particle in Mink$_4$ of boost weight $\delta$
\begin{equation}
\begin{split}
    \lim_{\ell \rightarrow \infty}\ell^{\Delta} \Psi_{\Delta}^{\rm null}&(\delta, \hat{q}(z); \tau, r, w)   \\
 &= \frac{\Gamma(\Delta - \delta)}{\Gamma(\Delta - \frac{1}{2})}  \frac{1}{(r + u - r\Omega \cdot \Omega_q + i\epsilon)^{\delta}}. 
    \end{split}
\end{equation}

\section{Discussion}

There are many future directions to be explored. The null foliation reveals perhaps unexpected similarities between quantum fields in AdS and flat spacetimes. These similarities should persist for gauge theories and gravity since, at the linearized level, the gauge field and graviton wavefunctions are related to the scalar ones by multiplication with polarization tensors \cite{Costa:2014kfa}. 
It will be interesting to revisit the AdS/CFT dictionary in this language, while also extending the analysis to interacting theories. Assuming that a basis- and quantization-independent version of the AdS/CFT dictionary exists, the principal series representations appearing in the null quantization should play a role in the story, which remains to be better understood. 

It will also be interesting to explore the implications of our reconstruction formula \eqref{eq:G-int} in the reconstruction of fields behind a black hole horizon \cite{Papadodimas:2012aq,deBoer:2022zps,Liu:2025krl,Ceplak:2025dds}. Our analysis suggests that the shadow transform should play an important role which remains to be elucidated. 

Since $\mathfrak{so}(3,1)$ is a subalgebra of the isometry algebras of all maximally symmetric spacetimes, we expect solutions to the Wheeler-DeWitt equations to share universal properties when expressed in the boost basis constructed here and its flat space and dS counterparts. A nice playground to explore this idea may be the low dimensional models  where these equations may admit exact solutions \cite{Anninos:2024iwf,Alonso-Monsalve:2024oii,Iliesiu:2024cnh,Held:2024rmg,Callebaut:2025zpc}. We hope to address at least some of these interesting questions in the near future.

\section{Acknowledgements}

We are grateful to Dionysios Anninos, Jan de Boer, Joydeep Chakravarty, Laurent Freidel, Damian Galante, Mait\'a Micol, Richard Myers, Kostas Skenderis, Tianli Wang and Yifei Zhao for helpful discussions. We also thank Dionysios Anninos, Jan de Boer for comments on a draft,  and especially Dominik Neuenfeld for careful reading and questions that led to several improvements of the presentation. A.R. was in part supported by the Simons Foundation through the Emmy Noether Fellows Program at Perimeter Institute (1034867, Dittrich).
We are grateful for the hospitality of Perimeter Institute where part of this work was carried out.
Research at Perimeter Institute is supported in part by the Government of Canada through the Department of Innovation, Science and Economic Development and by the Province of Ontario through
the Ministry of Colleges and Universities.

\pagebreak \widetext
\section*{Appendix}

\subsection{Wavefunction asymptotics in the null foliation}
\label{app:wavefunction-asy}

The large-$r$ behavior of the wavefunctions \eqref{eq:null-wavefunctions} is given by
\begin{equation}
\begin{split}
    R(r) &\sim r^{-\Delta} \Gamma\left(\frac{3}{2} - \Delta\right) \left(\frac{c_1 \Gamma(\frac{1}{2} - \ell)}{\Gamma(\frac{1}{2}(2 - \Delta - \ell \pm  \omega))} + \frac{c_2\Gamma(\frac{3}{2} + \ell)}{\Gamma(\frac{1}{2}(3 - \Delta + \ell \pm  \omega))}  \right)\\
    &+ r^{-3+\Delta} \Gamma\left(-\frac{3}{2} +\Delta\right)\left(\frac{c_1 \Gamma(\frac{1}{2} - \ell)}{\Gamma(\frac{1}{2}( \Delta -1 - \ell \pm \omega))} + \frac{c_2\Gamma(\frac{3}{2} + \ell)}{\Gamma(\frac{1}{2}(\Delta + \ell \pm \omega))}  \right) + \cdots ,
    \end{split}
\end{equation}
where we defined
\begin{equation}
    \Gamma(x \pm \omega) \equiv \Gamma(x + \omega) \Gamma(x -\omega).
\end{equation}
The growing mode at large-$r$ can be eliminated by setting
\begin{equation}
\label{eq:bc}
 c_1 = -\frac{c_2\Gamma(\frac{3}{2} + \ell) \Gamma(\frac{1}{2}( \Delta -1 - \ell \pm \omega))}{\Gamma(\frac{1}{2} - \ell)\Gamma(\frac{1}{2}(\Delta + \ell \pm \omega))} .
\end{equation}
For this choice, the small-$r$ behavior of \eqref{eq:null-wavefunctions} is 
\begin{equation}
    R(r) \sim c_2 \left( r^{\ell} - \frac{(-1)^{1-\ell}}{r^{\ell + 1}} \frac{ \Gamma\left(\frac{1}{2}(\Delta  - \ell - 1\pm\omega)\right)\Gamma(\ell + \frac{3}{2})}{\Gamma(\frac{1}{2} - \ell) \Gamma(\frac{1}{2}(\Delta + \ell\pm\omega))} \right),
\end{equation}
which diverges in as $r \rightarrow 0$ for any $\ell$. Conversely, demanding that the solutions are regular near the origin fixes $c_1 = 0$, in which case $R(r)$ will contain both normalizable and non-normalizable modes near the boundary for $\Delta \in \mathbb{R}$.

\bibliographystyle{bibstyle}
\bibliography{references}

\end{document}